\documentclass[12pt,english]{article}
\usepackage[latin9]{inputenc}
\usepackage{geometry}
\usepackage[active]{srcltx}
\usepackage{amsthm}
\usepackage{amsmath}
\usepackage[numbers]{natbib}

\makeatletter
\numberwithin{equation}{section}

\@ifundefined{date}{}{\date{}}
\makeatother

\usepackage{babel}
\begin{document}

\title{The Bayesian analysis of contingency table data using the bayesloglin
R package}

\author{Matthew Friedlander}

\maketitle

\section{Introduction}

Data in the form of a contingency table arise when individuals are
cross classified according to a finite number of criteria. Log-linear
modeling (see e.g., \citep{Agresti}, \citep{Christensen}, or \citep{Bishop})
is a popular and effective methodology for analyzing such data enabling
the practitioner to make inferences about dependencies between the
various criteria. For hierarchical log-linear models, the interactions
between the criteria can be represented in the form of a graph; the
vertices represent the criteria and the presence or absence of an
edge between two criteria indicates whether or not the two are conditionally
independent \citep{Lauritzen}. This kind of graphical summary greatly
facilitates the interpretation of a given model. 

For log-linear analysis, we can use the conjugate prior of \citep{Massam}
to work in the Bayesian paradigm. With this prior, the MC3 algorithm
of \citep{Madigan} allows for exploration of the space of models
to try to find those with the highest posterior probability. Once
top models have been identified, a block Gibbs sampler can be constructed
to sample from the posterior distribution and to estimate parameters
of interest. Our aim in this paper, is to introduce the bayesloglin
R package \citep{R} to carry out these tasks. 

The outline of this paper is as follows: In section 2, we develop
the notation for hierarchical log-linear models which is based on
\citep{Letac}. In section 3, we give the conjugate prior for hierarchical
models under Poisson sampling. In section 4, we describe the MC3 algorithm
for searching the spaces of hierarchical, graphical, and decomposable
models. Section 5 deals with Gibbs sampling and section 6 gives some
exact results for the normalizing constant and the mean and variance
of the log-linear parameters for decomposable models. In section 7
we illustrate the use of the bayesloglin package for analyzing the
often studied Czech autoworkers data from \citep{Edwards}.

\section{Preliminaries}

The notation in this section is adapted from \citep{Letac} with some
minor changes to move from multinomial to Poisson sampling. Let $V$
be a finite set of indices representing $|V|$ criteria. We assume
that the criterion labeled by $v\in V$ can take values in a finite
set $\mathcal{I}_{v}$. The resulting counts are gathered in a contingency
table such that 
\[
\mathit{\mathcal{I}=}\prod_{v\in V}\mathcal{I}_{v}
\]
is the set of cells $i=\left(i_{v},v\in V\right)$. The vector of
cell counts is denoted $n=\left(n(i),i\in\mathcal{I}\right)$ with
corresponding mean $m(i)=E(n)=\left(m(i),i\in\mathcal{I}\right)$.
For $D\subset V,$
\[
\mathcal{I}_{D}=\prod_{v\in D}\mathcal{I}_{v}
\]
is the set of cells $i_{D}=(i_{v},v\in D)$ in the $D$-marginal table.
The marginal counts are $n(i_{D})=\sum_{j:j_{D}=i_{D}}n(j)$ with
$m(i_{D})=E\left(n(i_{D})\right)$. 

Let $\mathcal{D}$ be a family of subsets of $V$ such that $D\in\mathcal{D}$
and $D_{1}\subset D$ implies that $D_{1}\in\mathcal{D}$. We will
assume that $\cup_{D\in\mathcal{D}}D=V$. The hierarchical log-linear
model generated by $\mathcal{D}$ is
\[
\log m(i)=\sum_{D\in\mathcal{D}}\lambda_{D}(i)
\]
where $m(i)$ is assumed positive and $\lambda_{D}(i)$ is a real
valued function that depends on $i$ only through $i_{D}$.

We now select a select a special element in each $\mathcal{I}_{v}$.
For convenience, we denote it $0$. We also denote $0$ in $i$ the
cell with all its components equal to $0$. The choice of special
element $0$ in each $\mathcal{I}_{v}$ is arbitrary. If $i\in\mathcal{I}$,
the support of $i$ is the subset of $V$ defined as $S(i)=\left\{ v\in V,i_{v}\ne0\right\} $.
We let $J=\left\{ j\in\mathcal{I},S(j)\in\mathcal{D}\right\} $ and
define the notation $j\triangleleft i$ for $i\in\mathcal{I}$ and
$j\in J$ to mean that $j_{S(j)}=i_{S(j)}$. By convention, we say
that $0\triangleleft i$ for any $i\in\mathcal{I}$. For any $a\in\mathcal{D}$,
we also define the sub-model $J_{a}=\left\{ j\in J:S(j)\subseteq a\right\} $.

Let $\left(e_{j},j\in J\right)$ be the canonical basis of $R^{J}.$
For all $i\in\mathcal{I},$ we define $f_{i}\in R^{J}$ by
\[
f_{i}=\sum_{j\in J,j\triangleleft i}e_{j}.
\]
The baseline constrained hierarchical log-linear model generated by
$\mathcal{D}$ has the unique representation
\[
\log m(i)=\sum_{j\in J:j\triangleleft i}\theta_{j}=\left\langle f_{i},\theta\right\rangle 
\]
for $i\in\mathcal{I}$ and $\theta=\left(\theta_{j},j\in J\right)\in R^{J}$.
In matrix notation, we have
\[
\log m=X\theta
\]
where $X$ is an $\mathcal{I}\times J$ design matrix of full column
rank with rows $\left\{ f_{i},i\in\mathcal{I}\right\} $. It is worth
noting that $X$ is a binary 0/1 matrix with a first column that is
all 1's.

\section{Prior distribution under Poisson sampling}

We assume that the components of $n$ are independent and follow a
Poisson distribution. The sufficient statistic $t=X^{T}n$ has a probability
distribution in the natural exponential family
\[
f(t)=\exp\left(\left\langle \theta,t\right\rangle -\sum_{i\in\mathcal{I}}\exp\left(\left\langle f_{i},\theta\right\rangle \right)\right)\nu\left(dt\right)
\]
with respect to a discrete measure $\nu$ that has convex support
\[
C^{p}=\left\{ \sum_{i\in\mathcal{I}}y(i)f_{i},y(i)\ge0,i\in\mathcal{I}\right\} =\mathrm{cone}\left\{ f_{i},i\in\mathcal{I}\right\} 
\]
i.e. the convex cone generated by the rows of the design matrix $X$.
The Diaconis and Ylvisaker \citep{Diaconis} conjugate prior with
respect to the Lebesgue measure for the log-linear parameters is 
\[
f(\theta)=I(r,\alpha)^{-1}\exp\left(\alpha\left\langle r,\theta\right\rangle -\alpha\sum_{i\in\mathcal{I}}\exp\left\langle f_{i},\theta\right\rangle \right)
\]
where 
\[
I(r,\alpha)=\intop_{\theta\in R^{J}}\exp\left(\alpha\left\langle r,\theta\right\rangle -\alpha\sum_{i\in\mathcal{I}}\exp\left\langle f_{i},\theta\right\rangle \right)d\theta
\]
and is proper when $\alpha>0$ and $r=X^{T}y$ for some $y>0$ i.e.
$r$ is in the relative interior of $C^{p}$. The Bayes factor for
comparing two models $J_{1}$ and $J_{2}$ is 
\[
B_{12}=\frac{P(t|J_{1})}{P(t|J_{2})}=\frac{I_{1}\left(\frac{t+\alpha r}{1+\alpha},1+\alpha\right)/I_{1}(r,\alpha)}{I_{2}\left(\frac{t+\alpha r}{1+\alpha},1+\alpha\right)/I_{2}\left(r,\alpha\right)}
\]

\section{The MC3 algorithm for model selection}

The Bayesian paradigm to model selection involves choosing models
with high posterior probability from a set $\mathcal{M}$ of competing
models. We associate with each model $J\in\mathcal{M}$ a neighbourhood
$\mathrm{nbd}\left(J\right)\subset\mathcal{M}$. The $MC^{3}$ algorithm
proposed by \citep{Madigan} constructs an irreducible Markov chain
with state space $\mathcal{M}$ and and equillibrium distribution
$\left\{ p(J|n):J\in\mathcal{M}\right\} $ where $P(J|t)$ is the
posterior probability of $J$. We assume that all models are apriori
equally likely; hence $P(J|t)$ is proportional to the marginal likelihood
$P(t|J)=I\left(t+r,\alpha+1\right)/I(r,\alpha)$.

If the chain is in state $J$ at we draw a candidate model $J'$ from
a uniform distribution on $\mathrm{nbd}(J)$. The chain moves to $J'$
with probability
\[
\mathrm{min}\left\{ 1,\frac{P\left(t|J\right)/\mathrm{\#nbd}(J)}{P(t|J')/\mathrm{\#nbd}(J')}\right\} 
\]
where $\#\mathrm{nbd}(J)$ denotes the number of neighbours of $J$.
Otherwise the chain does not move. The evaluation of the marginal
likelihoods and the specification of model neighbourhoods is done
with respect to the particular properties of the set of candidate
models considered. 
\begin{enumerate}
\item \textbf{Hierarchical log-linear models}. We calculate the marginal
likelihood through the Laplace approximation to the normalizing constants
for the prior and posterior distribution of the log-linear model parameters.
The neighbourhood of a hierarchical model $J$ consists of the hierarchical
models obtained from $J$ by adding one of its dual generators (i.e.
minimal terms not present in the model) or deleting one of its generators
(i.e. maximal terms present in the model). For details see \citep{Edwards}
and \citep{Dellaportas}.
\item \textbf{Graphical log-linear models}. We evaluate the marginal likelihood
using the Laplace approximation to the normalizing constants as we
do in the hierarchical case. The neighbourhood of a graphical model
with corresponding graph $G$ consists of those models whose independence
graphs are obtained from $G$ by adding or removing one edge. 
\item \textbf{Decomposable log-linear models}. In this case, the marginal
likelihood can be obtained explicitly. See Section 6 for the formula.
The neighbourhood of a decomposable model with corresponding graph
$G$ consists of those models whose independence graphs are decomposable
and are obtained by adding or deleting one edge from $G$.
\end{enumerate}

\section{Gibbs sampling}

Our aim in this section is to develop a blocked Gibbs sampler to sample
from the posterior distribution and to estimate parameters of interest.
We begin by partitioning the cells and the prior into blocks. For
$a\in\mathcal{D}$ we define the sets $B_{i_{a}}=\left\{ j\in\mathcal{I}:j_{a}=i_{a}\right\} $
for $i_{a}\in\mathcal{I}_{a}$. These sets are disjoint and partition
$\mathcal{I}.$ Define the vectors $\chi_{i_{a}},i_{a}\in\mathcal{I}_{a}$
with 
\[
\mathcal{\chi}_{i_{a}}(i)=\begin{cases}
1 & i\in B_{i_{a}}\\
0 & \mathrm{otherwise}
\end{cases}
\]
and the matrix $\chi$ with columns $x_{i_{a}}(i)$. We can then write:
$f_{i}=f_{(i_{a},i_{a}^{c})}=f_{(i_{a},0)}+f_{(0,i_{a}^{c})}$ and
$\theta=\left(\theta_{a},\theta_{a^{c}}\right)$ where $\theta_{a}=\left(\theta_{j}:S(j)\subseteq a\right)$
and $\theta_{a^{c}}=\left(\theta_{j}:S(j)\not\subset a\right)$.

The marginal counts $m(i_{a}),i_{a}\in\mathcal{I}_{a}$ follow a log-linear
model with 
\begin{eqnarray*}
m\left(i_{a}\right) & = & \sum_{i\in B_{i_{a}}}\exp\left(\left\langle f_{i},\theta\right\rangle \right)\\
 & = & \exp\left(\left\langle f_{(i_{a},0)},\theta\right\rangle \right)\sum_{i\in B_{i_{a}}}\exp\left(\left\langle f_{(0,i_{a^{c}})},\theta\right\rangle \right)
\end{eqnarray*}
and, taking logs,
\begin{eqnarray*}
\log m\left(i_{a}\right) & = & \sum_{i\in B_{i_{a}}}\left\langle f_{(i_{a},0)},\theta\right\rangle +\log\left(\sum_{i\in B_{i_{a}}}\exp\left(\left\langle f_{(0,i_{a^{c}})},\theta\right\rangle \right)\right)
\end{eqnarray*}
Let $m_{a}=\left(m(i_{a}),i_{a}\in\mathcal{I}_{a}\right)$ and partition
the matrix $X$ such that $X=\left[X_{a},X_{\bar{a}}\right]$ where
$X_{a}$ is a matrix made up of those columns of $X$ corresponding
to $j$ such that $S(j)\subseteq a$ and $X_{\bar{a}}$ is a matrix
with all the other columns. Then, in matrix notation, 
\[
\log m_{a}=\left(\frac{\chi^{T}X_{a}}{|\mathcal{I}\mathcal{\backslash I}_{a}|}\right)\theta_{a}+\log\left(\chi^{T}\exp\left(X_{\bar{a}}\theta_{\bar{a}}\right)\right)
\]
Returning to the prior, parametrized temporarily in terms of $m$,
we can partition $f$ as 
\begin{eqnarray*}
f(m|J) & \propto & \exp\left(\alpha\left\langle y,\log m\right\rangle -\alpha\sum_{i\in\mathcal{I}}m(i)\right)\\
 & = & \left\{ \prod_{i_{a}\in\mathcal{I}_{a}}\prod_{i\in B_{i_{a}}}\left(\frac{m(i)}{m\left(i_{a}\right)}\right)^{\alpha y(i)}\right\} \left\{ \prod_{i_{a}\in\mathcal{I}_{a}}m\left(i_{a}\right)^{\alpha y\left(i_{a}\right)}\exp\left(-\alpha m\left(i_{a}\right)\right)\right\} \\
 & = & f\left(m_{\bar{a}}\right)f\left(m_{a}|m_{\bar{a}}\right)
\end{eqnarray*}
and we see that $f\left(m_{a}|m_{\bar{a}}\right)$ is the product
of independent $\mathrm{Gamma}\left(1+\alpha y\left(i_{a}\right),1/\alpha\right),i_{a}\in\mathcal{I}_{a}$
distributions. Since it is easy to generate from $f\left(m_{a}|m_{\bar{a}}\right)$
for each $a\in\mathcal{D}$, a blocked Gibbs sampler \citep{Jensen}
is feasible to sample from $f(\theta)$. Following \citep{Dobra1},
we begin by choosing an arbitrary initial value of $\theta^{(0)}$.
For a given value of $\theta^{(k)}$, we update as follows: 
\begin{enumerate}
\item Generate independent $m\left(i_{a}\right)\sim\mathrm{Gamma}\left(\alpha y\left(i_{a}\right),1/\alpha\right)$
random variables for all $a\in\mathcal{D}$ and $i_{a}\in\mathcal{I}_{a}$.
\item For each $a\in\mathcal{D}$, in any arbitrary order set, 
\[
\theta_{a}^{(k)}=\left(\frac{\chi^{T}X_{a}}{|\mathcal{I}\mathcal{\backslash I}_{a}|}\right)^{-1}\left(\log\left(m_{a}\right)-\log\left(x^{T}\exp\left(X_{\bar{a}}\theta_{\bar{a}}\right)\right)\right)
\]
using the most recent value of $\theta_{\bar{a}}$ available.
\end{enumerate}
After a suitable burn-in, the resulting samples come from $f(\theta)$.
We note that the above Gibbs sampler is also known as the Bayesian
Iterative Proportional Fitting algorithm. See \citep{Dobra1},\citep{Gelman},\citep{Piccioni},
and\citep{Schafer} for more details.

\section{Some exact results for decomposable models}

For decomposable models, some exact results exist for the normalizing
constant and the mean and variance of the log-linear parameters. Let
us reconsider the prior defined in section 3 as

\[
f(\theta)=I(r,\alpha)^{-1}\exp\left(\alpha\left\langle r,\theta\right\rangle -\alpha\sum_{i\in\mathcal{I}}\exp\left\langle f_{i},\theta\right\rangle \right)
\]
with 
\[
I(r,\alpha)=\intop_{\theta\in R^{J}}\exp\left(\alpha\left\langle r,\theta\right\rangle -\alpha\sum_{i\in\mathcal{I}}\exp\left\langle f_{i},\theta\right\rangle \right)d\theta
\]
where $\alpha>0$ and $r=\left(r_{j},j\in J\right)\in\mathrm{ri\left(C_{p}\right)}$.
Then
\[
\mathrm{E}\left(\alpha\theta\right)=\frac{\partial\log I(r,\alpha)}{\partial r}
\]
and 
\[
\mathrm{Cov}(\alpha\theta)=\frac{\partial^{2}\log I(r,\alpha)}{\partial r^{2}}
\]
In the case of log-linear models where $m$ is Markov with respect
to a decomposable graph $G=(V,E)$, with vertex set $V$ and edge
set $E$, an explicit formula exists for $I(r,\alpha)$. Let $C$
denote the set of cliques and $S$ the set of minimal vertex separators.
For a given $s\in S$, let $V_{1},V_{2},...,V_{p}$ be the connected
components of the subgraph $G_{V\backslash s}$ and $q$ be the number
of $j=1,2,...,p$ such that $s$ is not a clique of $S\cup V_{j}$.
Then $\nu(s)=q-1$ is called the multiplicity of $s$ and $\sum_{s\in S}\nu(s)=|C|-1$
\citep{Lauritzen}. Based on proposition 4.2 of \citep{Massam}, adapted
to Poisson sampling, we have 
\[
I(r,\alpha)=\alpha^{-\alpha\sum_{i\in\mathcal{I}}y(i)}\frac{\prod_{c\in C}\prod_{i_{c}\in\mathcal{I}_{c}}\Gamma\left(\alpha y(i_{c})\right)}{\prod_{s\in S}\prod_{i_{s}\in\mathcal{I}_{s}}\left\{ \Gamma\left(\alpha y(i_{s})\right)\right\} ^{\nu(s)}}
\]
Taking logs and differentiating with respect to $r$ gives
\[
E(\theta)=-\frac{d\sum_{i\in\mathcal{I}}y(i)}{dr}\log\alpha+\sum_{c\in C}\sum_{i_{c}\in\mathcal{I}_{c}}\psi\left(\alpha y(i_{c})\right)\frac{dy\left(i_{c}\right)}{dr}-\sum_{s\in C}\sum_{i_{s}\in\mathcal{I}_{s}}\nu(s)\psi\left(\alpha y(i_{s})\right)\frac{dy\left(i_{s}\right)}{dr}
\]
where $\psi$ is the digamma function. Note that the derivatives in
the right hand side of $\mathrm{E}(\theta)$ are vectors. In particular,
$d\sum_{i\in\mathcal{I}}y(i)/dr=\left(1,0,...,0\right)^{T}$ since
$r_{0}=\sum_{i\in\mathcal{I}}y(i)$. Differentiating once more we
have
\[
\mathrm{Cov}\left(\theta\right)=\sum_{c\in C}\sum_{i_{c}\in\mathcal{I}_{c}}\psi_{1}\left(\alpha y(i_{c})\right)\frac{dy\left(i_{c}\right)}{dr}\frac{dy\left(i_{c}\right)}{dr}^{T}-\sum_{s\in S}\sum_{i_{s}\in\mathcal{I}_{s}}\nu(s)\psi_{1}\left(\alpha y(i_{s})\right)\frac{dy\left(i_{s}\right)}{dr}\frac{dy\left(i_{s}\right)}{dr}^{T}
\]
with $\psi_{1}$ being the trigamma function. We note that for decomposable
models, the subgraphs $G_{c},c\in C$ and $G_{s},s\in S$ are all
complete and we have a saturated model on those subgraphs. For $a\in C\cup S$,
it is easy to find $d\left(y(i_{a})\right)/dr,i_{a}\in\mathcal{I}_{a}$
by inverting the design matrix for the model $J_{a}$.

\section{The bayesloglin R package.}

The bayesloglin package includes the $2^{6}$ Czech autoworkers data
from \citep{Edwards}. This cross-classification of 1841 men gives
six potential risk-factors for coronary thrombosis: (a) smoking, (b)
strenuous mental work, (c) strenuous physical work, (d) systolic blood
pressure, (e) ratio of beta and alpha lipoproteins and (f) family
anamnesis of coronary heart disease. Currently, bayesloglin only allows
choice of the hyperparameter $\alpha$ and sets $y(i)=1/|\mathcal{I}|$
for each $i\in\mathcal{I}$. Consequently, $r_{0}=\sum_{i\in\mathcal{I}}y(i)=1$.
The required R code to search for the top decomposable, graphical,
and hierarchical log-linear models is: 

\begin{verbatim}
> data(czech)
> s1 <- MC3 (init = NULL, alpha = 1, iterations = 5000, replicates = 1, 
              data = czech, mode = "Decomposable")
> s2 <- MC3 (init = NULL, alpha = 1, iterations = 5000, replicates = 1, 
              data = czech, mode = "Graphical")
> s3 <- MC3 (init = NULL, alpha = 1, iterations = 5000, replicates = 1, 
              data = czech, mode = "Hierarchical")
\end{verbatim}

\noindent The top models in terms of posterior probability are

\begin{verbatim}
> head(s1, n = 5)
                  formula   logPostProb 
1    [a,c,e][b,c][d,e][f]   5271.975 
2  [a,c,e][a,d,e][b,c][f]   5271.103 
3    [a,c,e][a,d][b,c][f]   5271.077 
4 [a,c][b,c][b,e][d,e][f]   5270.549 
5  [a,c,e][b,c][b,f][d,e]   5270.394 

> head(s2, n = 5)
                         formula  logPostProb		
1      [a,c][a,d,e][b,c][b,e][f]  7122.398 
2   [a,c][a,e][b,c][b,e][d,e][f]  7121.580 
3    [a,c][a,d,e][b,c][b,e][b,f]  7121.374 
4   [a,c][a,d][a,e][b,c][b,e][f]  7120.683 
5 [a,c][a,e][b,c][b,e][b,f][d,e]  7120.556 

> head(s3, n = 4)
                                 formula  logPostProb		
1      [a,c][a,d][a,e][b,c][c,e][d,e][f]  7125.171 
2      [a,c][a,d][a,e][b,c][b,e][d,e][f]  7124.704
3 [a,c][a,d][a,e][b,c][b,e][c,e][d,e][f]  7124.229       	  
4    [a,c][a,d][a,e][b,c][b,f][c,e][d,e]  7124.147  
\end{verbatim}

These results match those obtained by the same methods in \citep{Massam}.
Consider the top hierarchical model $[a,c][a,d][a,e][b,c][c,e][d,e][f]$.
We can use the function $\mathrm{\mathtt{gibbsSampler}}$ to sample
from the posterior and obtain estimates of the mean and variances
of the log-linear parameters. We use a burn-in of 5000 iterations.

\begin{verbatim}
> formula <- freq ~ a*c + a*d + a*e + b*c + c*e + d*e + f 
> s <- gibbsSampler (formula, alpha = 1, data = czech, 
                     nSamples = 15000, verbose = T) 
> postMean <- colSums(s[5000:15000,]) / 10000 
> postCov <- cov(s[5000:15000,]) 
> postVar <- diag(postCov)
\end{verbatim}The values of $\mathrm{\mathtt{postMean}}$ and $\mathtt{\mathrm{\mathtt{postVar}}}$
are 

\begin{verbatim}
> postMean 
(Intercept)          a1         c1         b1            d1          e1
  3.0915633  -0.4150080   1.0199107   0.9010453  -0.2877865  -0.4890538
         f1       a1:c1       b1:c1       a1:d1       a1:e1       c1:e1   
 -1.8057132   0.5409632  -2.8017859  -0.3542662   0.4871123  -0.4479492
      d1:e1    
  0.3784125 

> postVar 
(Intercept)          a1          c1          b1           d1           e1 
0.006921940 0.008033988 0.008498167 0.005310040  0.005564232  0.008184625
         f1       a1:c1       b1:c1       a1:d1        a1:e1        c1:e1
0.004433024 0.009185728 0.015035403 0.009219168  0.009280780  0.009133959
      d1:e1  
0.009298324  
\end{verbatim}We now consider the decomposable model $[a,c,e][b,c][d,e][f]$. The
$\mathrm{\mathtt{findPostMean}}$ and $\mbox{\ensuremath{\mathtt{\mathtt{findPostCov}}}}$
functions can compute the posterior mean and covariance matrix, which
for decomposable models, is available in closed form. In R we have

\begin{verbatim}
> formula <- freq ~ a*c*e + b*c + d*e + f
> postMean <- findPostMean (formula, alpha = 1, data = czech)
> postCov <- findPostCov(formula, alpha = 1, data = czech)
> postVar <- diag(postCov)

> postMean
(Intercept)          b1          c1          a1          e1          d1  
 3.1561271    0.9002899   1.0149757  -0.5565110  -0.4621862  -0.4387784
        f1        b1:c1       a1:c1       a1:e1       c1:e1       d1:e1  
-1.8051306   -2.8012942   0.5494842   0.4645452  -0.4380842   0.3412027
  a1:c1:e1
-0.0194745  


> postVar 
(Intercept)          b1          c1          a1          e1          d1  
0.006563014 0.005252849 0.009530313 0.008807288 0.009375078 0.003956279
         f1       b1:c1       a1:c1       a1:e1       c1:e1       d1:e1
0.004478660 0.014932109 0.015834157 0.018016838 0.018531263 0.009099995 
   a1:c1:e1  
0.037264994 
\end{verbatim}The reader can verify that the Gibbs sampler gives a close approximation
to the exact values for this model.\\

\noindent \textbf{Acknowledgements.} The author is grateful to the
creators of the $\mathtt{igraph}$ R package \citep{igraph} which
was used extensively for manipulating graphs. Special thanks also
to Adrian Dobra, whose C++ code for representing hierarchical log-linear
models, was adapted for use in R. 

\bibliographystyle{plainnat}
\bibliography{mybib}

\end{document}